\documentclass[12pt]{article}
\begin{document}

\begin{center}
{\large \bf An application of the catastrophe theory to building the model of
elastic-plastic behaviour of materials.}
\par
{\large \bf Part 1. Uniaxial deformation (stress)}
\par
\bigskip L. N. Maurin, I. S. Tikhomirova
\footnote{e-mail: tihomir@ivanivo.ac.ru}
\par
{\sl Physics Department, Ivanovo State Univarsity
\par
Ermaka St. 39, Ivanovo, 153025, Russia}
\end{center}

\begin{abstract}
The uniaxial elastic-plastic deformation process is
considered. Mathematical model of this process was built. According to this
model all stable static states form the lattice, which is called the $\Delta
$-lattice.
\end{abstract}

The idea suggested in [1, 2] will be evolved into this and latter papers.
The idea consist in definition of variety of elastic-plastic states seen in
plastic deformation process. The catastrophe theory underlies this approach
to the problem. As it is known, the subject of the catastrophe theory is an
investigation of qualitative character of the equation solutions in
dependence on the parameters, which are presented in this equations. In this
article we study the elastic-plastic deformation from standpoint of
equilibrium.

There are various models in the plasticity theory, which are able to
describe plastic behavior of materials. But all these models are considering
materials with smooth deformation diagram (fig. 1a). Nevertheless there
exist broken deformation curves, which are typical for serrated yielding or
the Portevin Le-Chatelier effect (see figure 1b). The classical plasticity
theory has no mathematical models describing such a behaviour in deformation
process.


In this article we will try to build macroscopic model of elastic-plastic
behaviour of bodies under loading. We take into account the assumption of
existence of the equilibrium static states of the deforming material. These
equilibrium states are elastic states. Plastic states are not static
equilibrium. They are implemented in the transforming from one equilibrium
state (elastic state) to another. Thus we can construct the variety of
equilibrium states that is depicted on fig. 2 for ideal plastic materials.
As it is seen from fig. 2 the variety of stable states is the lattice which
we will call the $\Delta $-lattice, here stress is non-dimensional quantity,
i. e. $\sigma =\sigma / E$, where $E$ is the Young's modulus.


Let us consider the deformation process now. Firstly, the deformation of the
material is elastic (see the rod O$_{0}$A$_{0}$ on fig. 2). Hook's law
carries into effect, i.~e. $\sigma = \varepsilon $, where \textit{$%
\sigma $} is the non-dimensional stress, \textit{$\varepsilon $} is the
strain. The point A$_{0}$ is the last point of elastic states on the first
rod of the $\Delta $-lattice. The elastic branch O$_{0}$A$_{0}$ doesn't
exist over this point (i.~e. for~\textit{$\varepsilon $~>~$\varepsilon $}$%
_{A0})$. As soon as deformation \textit{$\varepsilon $} is grater then
\textit{$\varepsilon $}$_{A0}$ there will be a jump to another equilibrium
branch O$_{1}$A$_{1}$. Further, as the strain \textit{$\varepsilon $}
exceeds value \textit{$\varepsilon $}$_{A1}$ the new jump on the branch O$%
_{2}$A$_{2}$ will take place etc. Thus the plastic deformation process is
spasmodic, i.~e. increase of plastic deformation value is accomplished by
jumps and is not smooth.

As it was said before we wish to use the catastrophe theory to build the
mathematical model of elastic plastic deformation. Everyone knows that the
investigation of the potential function (or the function of state) underlies
the catastrophe theory. That is why we must construct this function. The
function of state have to be minimum for each stable state. Consequently the
function of state is minimum on the rods of $\Delta $-lattice. Since the
minima of smooth function are to be separated by maximum then unstable
states must exist between the rods of $\Delta $-lattice. That is why the $%
\Delta $-lattice depicted on fig. 2 is required in additional constructions.
They are the joining the upper end (A$_{i})$ of each rod with lower end (O$%
_{i+1})$ of the next rod of $\Delta $-lattice (dotted lines on fig. 3). The
function of state is maximum for each point of the dotted lines, i.~e. these
states are unstable. The unstable states form another lattice, which will be
called additional variety.


Everything we need now is done and we can build the function of state. But
before it we must discuss the question of parameters. That means that we
will discuss the possible geometry of jumps. At first we consider the model
with parameter \textit{$\sigma $}, i.~e. stress is externally changed in
experiments. In this case it can be realized only the following types of
jumps (see fig. 4). First one is an ``ideal'' jump, when the stress is
constant and equal \textit{$\sigma $}$^{\ast }$, elastic strain is constant
as well (see the change A$\rightarrow $B on fig. 4). We must emphasize that
the total strain is increasing. The second passage is the passage A$%
\rightarrow $B$_{1}$ (see fig. 4) with increasing stress \textit{$\sigma $}
and strain \textit{$\varepsilon $}. The jump with decreasing stress \textit{$%
\sigma $ } from point A can not be realized because when stress is decreased
the elastic unloading will be realized (branch AO on fig. 4).


Now we will consider experiments with external changing parameter \textit{$%
\varepsilon $} (hard loading device as an example). In this case the
``ideal'' jump is A$\to $B (see fig. 5) with constant strain \textit{$%
\varepsilon $} and decreasing stress \textit{$\sigma $}. Another possible
passage in the model with external parameter \textit{$\varepsilon $} is the
passage at increasing strain \textit{$\varepsilon $} and decreasing stress
\textit{$\sigma $} (see the passage A$\to $B$_{1}$ on fig. 5). Also we
suppose that the jump at constant stress and increasing strain (A$\to $B$%
_{2} $ at fig. 5) can be implemented under definite conditions.

Another type of passages, namely the passages at increasing stress and
strain (A$\rightarrow $B$_{3}$ on fig. 5) and the passages at decreasing
stress and strain, cannot be realized in the model with parameter \textit{$%
\varepsilon $}.

Now we will consider experiments with external changing parameter \textit{$%
\varepsilon $} (hard loading device as an example). In this case the
``ideal'' jump is A$\rightarrow $B (see fig. 5) with constant strain \textit{%
$\varepsilon $} and decreasing stress \textit{$\sigma $}. Another possible
passage in the model with external parameter \textit{$\varepsilon $} is the
passage at increasing strain \textit{$\varepsilon $} and decreasing stress
\textit{$\sigma $} (see the passage A$\rightarrow $B$_{1}$ on fig. 5). Also
we suppose that the jump at constant stress and increasing strain (A$%
\rightarrow $B$_{2}$ at fig. 5) can be implemented under definite conditions.

Another type of passages, namely the passages at increasing stress and
strain (A$\to $B$_{3}$ on fig. 5) and the passages at decreasing stress and
strain, cannot be realized in the model with parameter \textit{$\varepsilon $%
}.


In accordance with geometry of jumps offered above we can distinguish two
types of state functions. The first one is the state function with parameter
\textit{$\sigma $}, and the second one is the state function with parameter
\textit{$\varepsilon $}.

Let us summarize the demands that the state function have to satisfy. The
state function must be smooth and has minima on every rod of $\Delta $%
-lattice and maxima for rods of additional variety. In accordance with these
demands we start to construct the state functions. Firstly we consider the
state function with parameter \textit{$\sigma $}. The function looks like
this:

(1) \ \ \ \ \ \ \ \ \ \ \ \ \ \ \ \ \ \ \ \ \ \ \ \ \ \ \
\ \ \ \ $\Phi =\Phi _{n}(\varepsilon \,;\,\,\sigma )$,

(2a) $\ \ \ \ \ \ \ \ \ \ \ \ \ \ \ \ \ \ \ \Phi _{n}^{/}\equiv
\frac{d\Phi _{n}}{d\varepsilon }=-\prod\limits_{s\,=\,-k}^{k}{\left( {\sigma
-\varphi _{n-s}(\varepsilon )}\right) }$,

(2b) \ \ \ \ $\varepsilon _{i}\,\leq
\,\varepsilon \,\,<\,\,\varepsilon _{i+1}$ , if \textit{$\varepsilon $}$_{i}$
-- projection of lower end of rod onto \textit{$\varepsilon $} axis,

\ \ \ \ \ \ \ \ \ \ \textit{$\varepsilon $}$%
_{i+1}$ -- projection of lower end of next rod onto \textit{$\varepsilon $}
axis;

(2c) \ \ \ \  $\varepsilon _{i}\,\leq
\,\varepsilon \,\,\leq \,\,\varepsilon _{i+1}$ , if \textit{$\varepsilon $}$%
_{i}$ -- projection of lower end of rod onto \textit{$\varepsilon $} axis,

\ \ \ \ \ \ \ \ \ \ \textit{$\varepsilon $}$%
_{i+1}$ -- projection of upper end of rod onto \textit{$\varepsilon $} axis
that is defined
\par
\ \ \ \ \ \ \ \ \ \ by function $\varphi _{l}$, where $l<i$;

(2d) \ \ \ \ $\varepsilon _{i}\,<\,\,\varepsilon
\,\,<\,\,\varepsilon _{i+1}$ , if \textit{$\varepsilon $}$_{i}$ --
projection of upper end of rod onto \textit{$\varepsilon $}
\par
\ \ \ \ \ \ \ \ \ \ axis,

\ \ \ \ \ \ \ \ \ \ \textit{$\varepsilon $}$%
_{i+1}$ -- projection of lower end of next rod onto \textit{$\varepsilon $}
axis;

(2e) \ \ \ \ $\varepsilon _{i}\,<\,\,\varepsilon
\,\,\leq \,\,\varepsilon _{i+1}$ , if \textit{$\varepsilon $}$_{i}$ --
projection of upper end of rod onto \textit{$\varepsilon $}
\par
\ \ \ \ \ \ \ \ \ \ axis,

\ \ \ \ \ \ \ \ \ \ \textit{$\varepsilon $}$%
_{i+1}$ -- projection of upper end of next rod onto \textit{$\varepsilon $}
axis;

(3) \ \ \ \ \ $\Phi _{n}\left( {\varepsilon _{n+1}}\right)
=\Phi _{n+1}\left( {\varepsilon _{n+1}}\right) $-- lacing condition,
\par
\noindent
where $n<(N-1)$, $N$ -- number of intervals on \textit{$\varepsilon $} axis.

\noindent Here for rods of $\Delta $-lattice

(4) \ \ \ \ \ \ $i=2m$, $\varphi _{i}(\varepsilon )=\varepsilon
\,\,-\,\sum\limits_{l=0}^{\frac{i}{2}-1}{\Delta _{l}}$,

\noindent and for rods of additional variety

(5) $\ \ \ \ \ \ i=2\cdot m+1$, $\ \ \ \ \varphi _{i}(\varepsilon
)=\,\,k_{i}\cdot \left( {\varepsilon \,\,-\,\sum\limits_{l=0}^{\frac{i-1}{2}}%
{\Delta _{l}}}\right) $,

\noindent where $k_i = \frac{\left( {\varepsilon - \sum\limits_{i = 0}^{m -
1} {\Delta _l } } \right)}{\left( {\varepsilon - \sum\limits_{i = 0}^m {%
\Delta _l } } \right)} = 1 + \frac{\Delta _m }{\varepsilon - \sum\limits_{i
= 0}^m {\Delta _l } }$, and \textit{$\varepsilon $} is equal to the strain
in point A$_{m}$.

As it is seen from equations (1) -- (3) the state function is constructed of
all $\varphi \,_{i}\left( \varepsilon \right) $ existing on the examining
part of the strain axis \textit{$\varepsilon $}. It is not difficult to test
(using (1)~--~(5)) that the state function $\Phi _{n}$ is minimum on all
rods of $\Delta $-lattice (i.~e. these states are stable) and maximum on all
rods of additional variety (unstable states). It is necessary to notice that
the end points of rods are degenerated critical points.

The second state function type is the state function with parameter \textit{$%
\varepsilon $}. We define this function as:

\textbf{(}6\textbf{) \ \ \ \ \
\ \ }$\Phi \,\,=\,\,\Phi _{n}\left( {\sigma \,;\,\,\varepsilon }%
\right) $\textbf{, } \ \ \ \ $\Phi _{n}(\sigma _{n+1})\,=\,\Phi _{n+1}(\sigma _{n+1})$%
\textbf{,}

(7) $\ \ \ \ \ \Phi _{n}^{/}=\,\,%
\frac{d\Phi _{n}}{d\sigma }\,\,\,=\,\,-\,\prod\limits_{i\,=\,2n}^{2(N\,-\,1)}%
{\left( {\varepsilon \,\,-\,\psi _{i}(\sigma )}\right) }\,\,\,\,=\,\prod%
\limits_{i\,=\,2n}^{2(N\,-\,1)}{\left( {\,\psi _{i}(\sigma )\,-\,\varepsilon
}\right) }$,

\noindent defined for region $\sigma _n \, < \,\sigma \,\, \le \,\sigma _{n
+ 1} $, if $n\, \ne \,0$,

\noindent and for $\sigma _n \,\, \le \,\,\sigma \,\, \le \,\sigma _{n + 1} $%
, if $n\, = \,0$, here N stands for number of rods of $\Delta $-lattice.

\bigskip

\noindent And $\psi _{i}(\sigma )=\sigma \,\,+\,\sum\limits_{l=0}^{\frac{i}{2%
}-1}{\Delta _{l}}$, for the rods of $\Delta $-lattice, where $i=2m$,

\noindent for the rods of additional variety $\psi _{i}(\sigma )=\,\,\frac{1%
}{k_{i}}\cdot \sigma \,\,\,+\,\sum\limits_{l=0}^{\frac{i-1}{2}}{\Delta _{l}}$%
, where $i=2\cdot m+1$.

Here $k_{i}$ is the same as in the case of a model with parameter \textit{$%
\sigma $}.

Like the previous state function (with parameter \textit{$\sigma $}) this
one is formed by means of all functions $\psi _i \left( \sigma \right)$
existing on the examining part of the stress axis \textit{$\sigma $}, i.~e.
it is formed by means of all rods of $\Delta $-lattice and additional
variety existing on the examining part of the \textit{$\sigma $}-axis. It is
easy to verify that the state function $\Phi _n \left( {\sigma
\,;\,\,\varepsilon } \right)$ (see (6)~--~(7)) also satisfies the above
demands.

In conclusion we emphasize that in accordance with our point of view the
continuous deformation curve must be replaced by $\Delta $-lattice (fig. 6).
Only the points of $\Delta $-lattice are the static stable states here.
There are no static stable states between the rods of $\Delta $-lattice (for
instance, points C, D, E, F, J on fig. 6).


So the state function was build for the macroscopic models of
elastic-plastic deformation both with parameters \textit{$\varepsilon $} and
\textit{$\sigma $}.

\small
\bigskip
\begin{center}
{\bf References}
\end{center}
\medskip
\small
\begin{itemize}
\item[[1{]}]
Maurin L. N. The lattice as the form of multitude of elastic-plastic states
of medium. Part 1. Macroscopic quantiztion of plasticity in the uniaxial
deformation. Vestnic Ivanovsk. Gosudarstven.
Universiteta. Seria Biologia, chimia, fisika, matematica. 2000.
Vip. 3. C. 76-82  (Russian)

\item[[2{]}]
Maurin L. N., Tikhomirova I. S. The lattice as the form of multitude of
elastic-plastic states of medium. Part 2. The model containing
the similarity hypothesis. Vestnic Ivanovsk. Gosudarstven.
Universiteta. Seria Biologia, chimia, fisika, matematica. 2000.
Vip. 3. C. 82-84  (Russian)

\end{itemize}

\newpage
\textbf{Fig. 1} \textbf{a, b: }a -- smooth deformation curve;
b -- deformation
curves \textit{$\sigma (\varepsilon )$} of pressing of nanocrystal (1) and
monocrystal (2) niobium.

\textbf{Fig. 2.} The $\Delta $-lattice of ideal plastic body.

\textbf{Fig. 3.} The $\Delta $-lattice and additional variety of ideal
plastic material.

\textbf{Fig. 4.} The possible transitions in the model with parameter
\textit{$\sigma $}.

\textbf{Fig. 5.} Possible (AB, AB$_{1}$, AB$_{2})$ and impossible (AB$_{3})$%
\ transitions in the model with parameter \textit{$\varepsilon $}.

\textbf{Fig. 6.} Replacement of continuous deformation curve by $\Delta $%
-lattice.

\end{document}